\documentclass[aps,showpacs,preprintnumbers,amsmath,amssymb,twocolumn]{revtex4}
\usepackage[dvips]{epsfig}
\usepackage{graphicx}
\usepackage{amssymb}
\usepackage{color}
\usepackage{enumerate}

\begin{document}
\baselineskip=0.4 cm
\title{\bf Double shadow of a regular phantom black hole as photons couple to Weyl tensor}

\author{Yang Huang, Songbai Chen$^{1}$\footnote{Corresponding author: csb3752@hunnu.edu.cn}, Jiliang
Jing$^{1}$ \footnote{jljing@hunnu.edu.cn}}

\affiliation{Institute of Physics and Department of Physics, Hunan
Normal University,  Changsha, Hunan 410081, People's Republic of
China \\ Key Laboratory of Low Dimensional Quantum Structures \\
and Quantum Control of Ministry of Education, Hunan Normal
University, Changsha, Hunan 410081, People's Republic of China\\
Synergetic Innovation Center for Quantum Effects and Applications,
Hunan Normal University, Changsha, Hunan 410081, People's Republic
of China}

\begin{abstract}
\baselineskip=0.4 cm
\begin{center}
{\bf Abstract}
\end{center}

We have studied the shadow of a regular phantom black hole as photons couple to Weyl tensor. We find that the coupling yields that photons with different polarization directions propagate along different paths in the spacetime so that there exits double shadow for a black hole, which is quite different from that in the non-coupling case where only a single shadow emerges. The umbra of black hole increases with the phantom charge and decreases with the coupling strength. The dependence of the penumbra on the phantom charge and the coupling strength is converse to that of the umbra. Combining with the supermassive central object in our Galaxy, we estimated the shadow of the black hole as the photons couple to Weyl tensor. Our results show that the coupling brings richer behaviors of the propagation of coupled photon and the shadow of the black hole in the regular phantom black hole spacetime.

\end{abstract}

\pacs{ 04.70.Dy, 95.30.Sf, 97.60.Lf } \maketitle

\section{Introduction}

The increasing evidences support that there exist supermassive black holes at
the center of many galaxies. Thus, detecting black hole parameters becomes very important since it can help us to identify black hole and to understand further features of black hole. Recent investigations show that the shadow of black hole carries the information about the black hole since the shape and size of the shadow are determined by black hole parameters \cite{sha1,sha2,sha3}. Compared with relativistic images in strong gravitational lensing, it is widely believed that the shadow of a black hole could be easily observed since it is a two-dimensional dark zone on the observer's sky. In general, the shadow for a static black hole is a perfect circle \cite{sha1}. For the rotating case, the shadow
has an elongated shape in the direction of the rotation axis
due to the dragging effect \cite{sha2,sha3}. Motivated by that the investigation of the shadow is very useful for measuring the nature of the black hole and the corresponding observations may be obtained in the near future, a lot of attention have been attracted on this subject in the last few years \cite{sha4,sha5,sha6,sha7,sha9,sha10,sha11,sha12,sha13,sha14,sha15,sha16,sha17}.

It is well known that the shadow of black hole is determined by the propagation of light ray in the spacetime, which depends on the parameters of background black hole, the properties of light itself and the interactions between light and other fields. Since light is actually a
kind of electromagnetic wave, it means that the interactions between Maxwell tensor and Weyl tensor will affect the propagation of photons in the black hole spacetime and result in some particular optical phenomena. The coupling between Maxwell tensor and Weyl tensor is investigated firstly in \cite{Drummond} by considering the effects of one-loop vacuum polarization on the photon effective action for quantum electrodynamics. Especially, the coupling arising from quantum corrections changes both the path and the maximum  velocity of photon propagation, which could result in  the ``superluminal" phenomenon in some cases \cite{Drummond,Daniels,Caip,Cho1,Lorenci}. Moreover, the influence of the vacuum polarisation on the propagation of the low-energy electromagnetic radiation have been analysed in the black hole spacetime \cite{Emelyanov}.
However, as a quantum phenomenon, the strength of the effects is immeasurably small in this low energy effective theory since the coupling constants are very small and are of the order of the square of the Compton wave length of the electron. Recently, the extended theoretical models without the small coupling constant limit have been investigated
for some physical motivations \cite{Turner,Bamba,Ni,Solanki,Dereli1,Balakin,Hehl} .
The optical behaviors of the coupled photons in the extended Weyl correction model have been studied in the strong field region \cite{sb0,xy}, which shows that the measurement of relativistic images and time delay in the strong field can provide a mechanism to detect the polarization direction of coupled photons. However, to my knowledge, what effects of the coupling between photon and Weyl tensor on the shadow of black hole have not been studied elsewhere, even for Schwarzschild black hole.

On the other hand, phantom dark energy is an exotic kind of theoretical models with the negative kinetic energy\cite{Caldwell}. Though it can interpret the accelerating expansion of the current Universe \cite{ph1,ph2,ph3,ph4,ph5,ph6}, it is still a big challenge for physics to under the strange effects caused by phantom dark energy including the violation of the null energy condition, the big rip of the Universe dominated by phantom, the valid of the cosmic censorship conjecture, and so on. Especially,  phantom dark energy models are still favoured by many recent precise observational data \cite{ph7}.
Thus, it is necessary to investigate further phantom field in various fields of physics. Recently, Bronnikov \textit{et al} \cite{pbh3} obtained a kind of solutions with phantom scalar hair, which has the properties of both black holes and wormholes. The stability of such a solution supported by a scalar field with a negative kinetic term has been investigated  \cite{Bronnikov:2012ch}. Some other black hole solutions describing gravity coupled to phantom scalar fields or phantom Maxwell fields have been found  and the corresponding geometric structure and thermodynamic properties are also studied in \cite{pbh1,pbh2,pbh4,pbh5,pbh6,pbh8,pbh9}. The strong gravitational lensing of such kind of black holes with phantom hair has been investigated in \cite{pbh7,GL1,GL2,GL3,GL4}. In this paper, we are going to study shadow of a regular phantom black hole \cite{pbh3} as photons couple to Weyl tensor and then probe what effects of the phantom charge of black hole, the photon polarization directions and the coupling between photon and Weyl tensor on the properties of the shadow. Moreover, we can obtain the such kind of effects on Schwarzschild black hole through setting phantom charge equal to zero since the usual Schwarzschild black hole is only a special case of the phantom black hole without phantom charge, which is another reason why we here chose such kind of phantom black holes as a background.

The paper is organized as follows: In section II, we
present equation of motion for the photons coupled to
Weyl tensor in the regular phantom black hole spacetime
and study the effects of the coupling on the inner
circular orbit radius of the photon around the black
hole. In section III, we will investigate the dependence
of the double shadow on the phantom charge
of black hole, the photon polarization directions and
the coupling between photon and Weyl tensor. Finally,
we end the paper with a summary.

\section{Equation of motion for the photons coupled to Weyl tensor and inner circular orbit radius}

Let us now review in brief the regular and static phantom black hole obtained by Bronnikov \textit{et al} in \cite{pbh3}. The action with phantom scalar field $\Phi$ in the curve spacetime is
\begin{eqnarray}
S=\int \sqrt{-g}d^4x[R-\frac{1}{2}g^{\mu\nu}\partial_{\mu}\Phi\partial_{\nu}\Phi+V(\Phi)].\label{act1}
\end{eqnarray}
It is found that this action admits a solution describing the gravity of a regular and static spacetime with phantom scalar hair, whose metric has a form \cite{pbh3}
\begin{eqnarray}
ds^2&=&-f(r)dt^2+\frac{1}{f(r)}dr^2+(r^2+b^2)\times\nonumber\\&&(d\theta^2+r^2\sin^2{\theta}
d\phi^2),\label{metric0}
\end{eqnarray}
with
\begin{eqnarray}
f(r)=1-\frac{3M}{b}\bigg[(\frac{\pi}{2}-\arctan\frac{r}{b})
(1+\frac{r^2}{b^2})-\frac{r}{b}\bigg].\label{f0}
\end{eqnarray}
The scalar field and the potential are
$\Phi\equiv\sqrt{2}\psi=\sqrt{2} \arctan \frac{r}{b}$ and
$V=\frac{3M}{b^3}[(\frac{\pi}{2}-\psi)(3-2\cos^2\psi)-3\sin\psi\cos\psi]$, respectively.
Here $M$ is the Schwarzschild mass defined in the usual
way and $b$ is a positive constant related to the charge of phantom scalar field. The presence of phantom hair bring richer properties for the spacetime (\ref{metric0}). The radius of event horizon $r_H$ is in the range $0<r_H<2M$ as $0<b<\frac{3\pi M}{2}$ and becomes $r_H=0$ as $b=\frac{3\pi M}{2}$. When $b>\frac{3\pi M}{2}$, there doesn't exist any event horizon since the value of $r_H$  becomes negative, and then a throat appears like in wormholes.
Thus, this regular phantom solution has the properties of both black holes and wormholes \cite{pbh3,GL2}. Moreover, it is shown that as $0<b<\frac{3\pi M}{2}$, instead of a singularity, an expanding and asymptotically de Sitter
Kantowski-Sachs cosmology occurs in its internal region $r<r_H$ \cite{pbh3,pbh8}. This particular type of black holes have also been named ``black universes'', which have an interesting cosmological behavior in their internal region \cite{pbh31}. When $b$ tends to zero, it is easy to find that the phantom scalar field $\Phi$ becomes a constant $\frac{\sqrt{2}\pi}{2}$ and the corresponding potential $V$ approaches to zero, which yields that the action turns to the usual action without any material field and then the corresponding black hole metric (\ref{metric0}) reduces that of the usual Schwarzschild black hole one.

The Lagrangian density of the electromagnetic field coupling to Weyl
tensor $C_{\mu\nu\rho\sigma}$ can be expressed as \cite{Weyl1}
\begin{eqnarray}
L_{em}=-\frac{1}{4}F_{\mu\nu}F^{\mu\nu}+\alpha
C^{\mu\nu\rho\sigma}F_{\mu\nu}F_{\rho\sigma},\label{acts}
\end{eqnarray}
where $F_{\mu\nu}$ and $\alpha$ are the usual electromagnetic tensor and the coupling constant with dimension of length-squared, respectively. In a four-dimensional spacetime with metric $g_{\mu\nu}$, Weyl tensor $C_{\mu\nu\rho\sigma}$ is defined as
$C_{\mu\nu\rho\sigma}=R_{\mu\nu\rho\sigma}-(g_{\mu[\rho}R_{\sigma]\nu}-g_{\nu[\rho}R_{\sigma]\mu})+\frac{1}{3}R
g_{\mu[\rho}g_{\sigma]\nu}$, where the brackets around indices refers to the antisymmetric part. The Maxwell equation is modified as
\begin{eqnarray}
\nabla_{\mu}\bigg(F^{\mu\nu}-4\alpha
C^{\mu\nu\rho\sigma}F_{\rho\sigma}\bigg)=0.\label{WE}
\end{eqnarray}
From the above corrected Maxwell equation (\ref{WE}), one can obtain the equation of motion of the coupled photons by the geometric optics approximation in which the wavelength $\lambda$ of photon is much smaller than a typical curvature scale $L$, but is larger than the electron Compton wavelength $\lambda_c$, i.e., $\lambda_c<\lambda<L$. Under this approximation, the electromagnetic field strength can be written as
\begin{eqnarray}
F_{\mu\nu}=f_{\mu\nu}e^{i\theta},\label{ef1}
\end{eqnarray}
where $f_{\mu\nu}$ is a slowly varying amplitude so that the derivative term $f_{\mu\nu;\lambda}$ can be neglected in this approximation \cite{Drummond,Daniels,Caip,Cho1,Lorenci}. The quantity $\theta$ is a rapidly varying phase and the wave vector is defined as $k_{\mu}=\partial_{\mu}\theta$, which can be treated as the coupled photon momentum as in the usual theory of quantum particle.
According to the Bianchi identity,
one can find that the amplitude $f_{\mu\nu}$ has a form $f_{\mu\nu}=k_{\mu}a_{\nu}-k_{\nu}a_{\mu}$,
where $a_{\mu}$ is the polarization vector satisfying the condition that
$k_{\mu}a^{\mu}=0$. Substituting Eq.(\ref{ef1})
into Eq.(\ref{WE}) and using the relationship above, we can get the equation of motion of photon coupling to Weyl tensor
\begin{eqnarray}
k_{\mu}k^{\mu}a^{\nu}+8\alpha
C^{\mu\nu\rho\sigma}k_{\sigma}k_{\mu}a_{\rho}=0.\label{WE2}
\end{eqnarray}
As it is expected,  the coupling between Weyl tensor and electromagnetic field changes the propagation of the coupled photon in the background spacetime.

For a regular phantom black hole spacetime (\ref{metric0}), introducing the vierbein fields
\begin{eqnarray}
e^a_{\mu}=diag(\sqrt{f},\;\frac{1}{\sqrt{f}},\;\sqrt{r^2+b^2},\;\sqrt{r^2+b^2}\sin\theta),
\end{eqnarray}
the metric $g_{\mu\nu}$ can be rewritten as
$g_{\mu\nu}=\eta_{ab}e^a_{\mu}e^b_{\nu}$, where $\eta_{ab}$ is the Minkowski metric. With the antisymmetric combination of
vierbeins defined in \cite{Drummond,Daniels}
\begin{eqnarray}
U^{ab}_{\mu\nu}=e^a_{\mu}e^b_{\nu}-e^a_{\nu}e^b_{\mu},
\end{eqnarray}
Weyl tensor can be simplified as
\begin{eqnarray}
C_{\mu\nu\rho\sigma}&=&\mathcal{A}\bigg(2U^{01}_{\mu\nu}U^{01}_{\rho\sigma}-
U^{02}_{\mu\nu}U^{02}_{\rho\sigma}-U^{03}_{\mu\nu}U^{03}_{\rho\sigma}
\nonumber\\&+&U^{12}_{\mu\nu}U^{12}_{\rho\sigma}+U^{13}_{\mu\nu}U^{13}_{\rho\sigma}-
2U^{23}_{\mu\nu}U^{23}_{\rho\sigma}\bigg),
\end{eqnarray}
with
\begin{eqnarray}
\mathcal{A}=-\frac{3Mr+b^2}{3(r^2+b^2)^2}.
\end{eqnarray}
Introducing three linear combinations of momentum components \cite{Drummond,Daniels}
\begin{eqnarray}
l_{\nu}=k^{\mu}U^{01}_{\mu\nu},\;\;\;\;
n_{\nu}=k^{\mu}U^{02}_{\mu\nu},\;\;\;\;
m_{\nu}=k^{\mu}U^{23}_{\mu\nu},
\end{eqnarray}
together with the dependent combinations
\begin{eqnarray}
&&p_{\nu}=k^{\mu}U^{12}_{\mu\nu}=\frac{1}{k^0}\bigg(k^1n_{\nu}-k^2l_{\nu}\bigg),\nonumber\\
&&r_{\nu}=k^{\mu}U^{03}_{\mu\nu}=\frac{1}{k^2}\bigg(k^0m_{\nu}+k^3l_{\nu}\bigg),\nonumber\\
&&q_{\nu}=k^{\mu}U^{13}_{\mu\nu}=\frac{k^1}{k^0}m_{\nu}+
\frac{k^1k^3}{k^2k^0}n_{\nu}-\frac{k^3}{k^0}l_{\nu},\label{vect3}
\end{eqnarray}
the equation of motion of the coupled photon (\ref{WE2}) can be simplified further as a set of equations for three independent polarisation components $a\cdot l$, $a\cdot n$, and $a\cdot m$,
\begin{eqnarray}
\bigg(\begin{array}{ccc}
K_{11}&0&0\\
K_{21}&K_{22}&
K_{23}\\
0&0&K_{33}
\end{array}\bigg)
\bigg(\begin{array}{c}
a \cdot l\\
a \cdot n
\\
a \cdot m
\end{array}\bigg)=0,\label{Kk}
\end{eqnarray}
with the coefficients
\begin{eqnarray}
K_{11}&=&(1+16\alpha \mathcal{A})(g^{00}k_0k_0+g^{11}k_1k_1)\nonumber\\&+&(1-8\alpha \mathcal{A})(g^{22}k_2k_2+g^{33}k_3k_3),\nonumber\\
K_{22}&=&(1-8\alpha \mathcal{A})(g^{00}k_0k_0+g^{11}k_1k_1\nonumber\\&+&g^{22}k_2k_2+g^{33}k_3k_3),
\nonumber\\
K_{21}&=&24\alpha \mathcal{A} \sqrt{g^{11}g^{22}}k_1k_2,\nonumber\\
K_{23}&=&24\alpha \mathcal{A}\sqrt{-g^{00}g^{33}}k_0k_3,\nonumber\\
K_{33}&=&(1-8\alpha \mathcal{A})(g^{00}k_0k_0+g^{11}k_1k_1)\nonumber\\&+&(1+16\alpha \mathcal{A})(g^{22}k_2k_2+g^{33}k_3k_3).
\end{eqnarray}
The non-zero solution of Eq.(\ref{Kk}) satisfies the condition $K_{11}K_{22}K_{33}=0$. The first root $K_{11}=0$ leads to the modified light cone
\begin{eqnarray}
&&(1+16\alpha \mathcal{A})(g^{00}k_0k_0+g^{11}k_1k_1)+\nonumber\\&&(1-8\alpha \mathcal{A})(g^{22}k_2k_2+g^{33}k_3k_3)=0, \label{Kk31}
\end{eqnarray}
which corresponds to the case the polarisation vector $a_{\mu}$ is proportional to $l_{\mu}$.  The second root corresponds to an unphysical polarisation and should be neglected. The third root is $K_{33}=0$, i.e.,
\begin{eqnarray}
&&(1-8\alpha \mathcal{A})(g^{00}k_0k_0+g^{11}k_1k_1)+\nonumber\\&&(1+16\alpha \mathcal{A})(g^{22}k_2k_2+g^{33}k_3k_3)=0,\label{Kk32}
\end{eqnarray}
which means that the vector $a_{\mu}=\lambda m_{\mu}$.  It is easy to find from Eqs.(\ref{Kk31}) and (\ref{Kk32}) that the effects of Weyl tensor on the photon propagation are
 different for the coupled photons with different polarizations, which leads to a phenomenon of birefringence. Moreover, the light cone conditions (\ref{Kk31}) and (\ref{Kk32})
imply that the motion of the coupled photons is non-geodesic in the original metric (\ref{metric0}). However, these photons follow null geodesics of the effective metric $\gamma_{\mu\nu}$ , i.e., $\gamma^{\mu\nu}k_{\mu}k_{\nu}=0$  [42]. The effective metric for the
coupled photon can be expressed as
\begin{eqnarray}
ds^2&=&-A(r)dt^2+B(r)dr^2+\nonumber\\
&&C(r)W(r)^{-1}(d\theta^2+\sin^2\theta d\phi^2),\label{l1}
\end{eqnarray}
where $A(r)=B(r)^{-1}=1-\frac{3M}{b}[(\frac{\pi}{2}-\arctan\frac{r}{b})
(1+\frac{r^2}{b^2})-\frac{r}{b}]$ and $C(r)=r^2+b^2$. The quantity $W(r)$ is
\begin{eqnarray}
W(r)=\frac{3(r^2+b^2)^2-8\alpha (b^2+3Mr)}{3(r^2+b^2)^2+16\alpha (b^2+3Mr)},
\end{eqnarray}
for photon with the polarization along $l_{\mu}$ (PPL ) and is
\begin{eqnarray}
W(r)=\frac{3(r^2+b^2)^2+16\alpha (b^2+3Mr)}{3(r^2+b^2)^2-8\alpha (b^2+3Mr)},
\end{eqnarray}
for photon with the polarization along $m_{\mu}$ (PPM ).

For the spherically symmetric metric (\ref{metric0}), we may consider only that the whole trajectory of the photon is limited on the equatorial plane $\theta=\frac{\pi}{2}$. Due to existence of cyclic coordinates $t$ and $\phi$ in spacetime (\ref{l1}), one can obtain the energy $E$ and angular momentum $L$ of the coupled photon as follow
\begin{eqnarray}
E=A(r)\dot{t},\;\;\;\;\;\;\;\;
L=C(r)W(r)^{-1}\dot{\phi},
\end{eqnarray}
where a dot represents a derivative with respect to affine parameter
$\lambda$ along the geodesics.  Making use of the relationship $k^{\mu}=\frac{dx^{\mu}}{d\lambda}$, one can
find that the equations of motion of coupled photon can be simplified further as
\begin{eqnarray}
\bigg(\frac{dr}{d\lambda}\bigg)^2
=\frac{1}{B(r)}\bigg[\frac{E^2}{A(r)}-W(r)\frac{L^2}{C(r)}\bigg].\label{v1}
\end{eqnarray}
The inner circular orbit radius $r_{ph}$ in the equatorial plane satisfied the conditions
\begin{eqnarray}
&&W(r)[A'(r)C(r)-A(r)C'(r)]\nonumber\\&&+A(r)C(r)W'(r)=0,\label{sp}
\end{eqnarray}
Here we set $E=1$. The changes of $r_{ph}$ with the coupling factor $\alpha$ and phantom charge $b$ for PPL and PPM  are plotted in Figs.(1) and (2), respectively.
\begin{figure}[ht]
\begin{center}
\includegraphics[width=3.8cm]{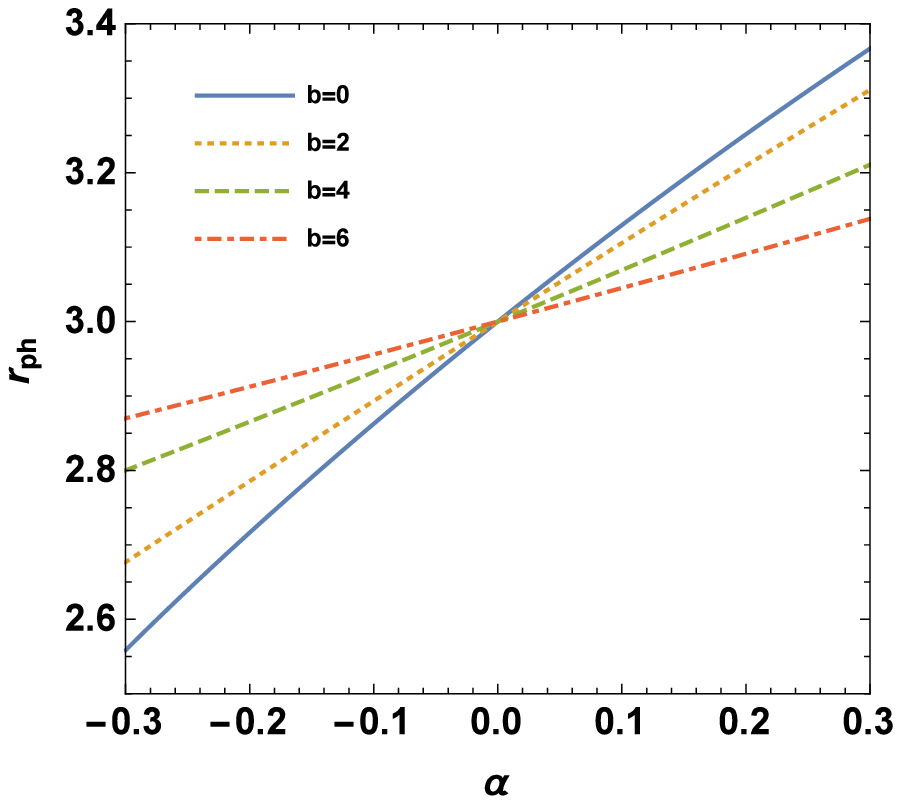}\;\;\;\includegraphics[width=3.8cm]{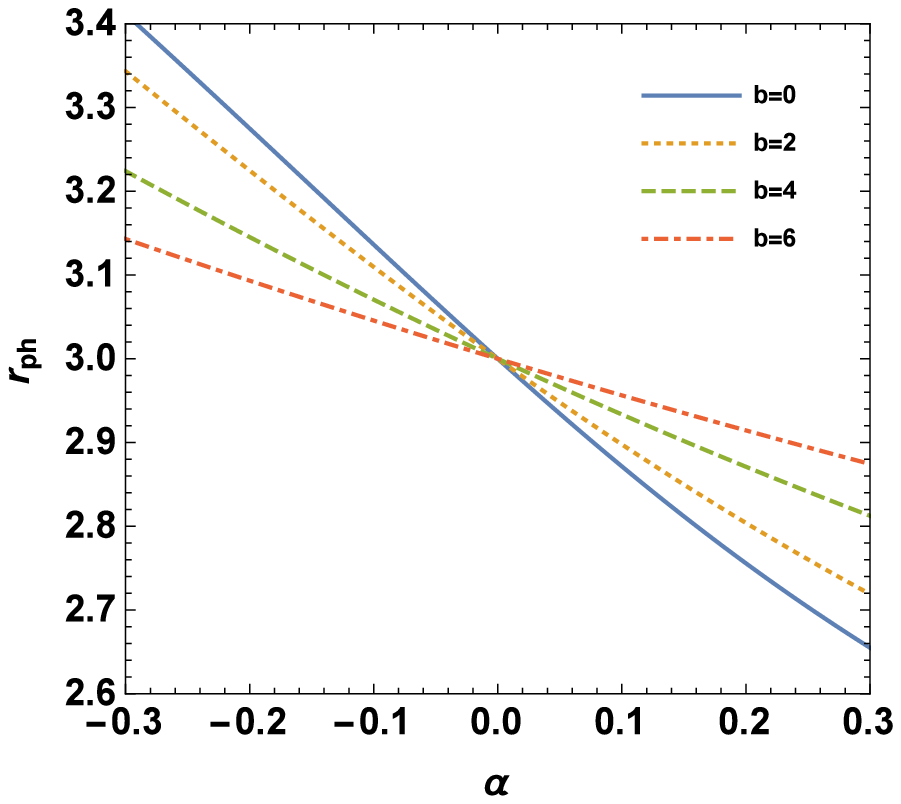}
\caption{Variety of the inner circular orbit radius $r_{ph}$ with
the coupling constant $\alpha$ in the regular phantom black hole
spacetime for fixed $b$. The left and the right are for the photons PPL and PPM,
respectively. Here we set $2M=1$.}
\end{center}
\end{figure}
\begin{figure}
\begin{center}
\includegraphics[width=3.8cm]{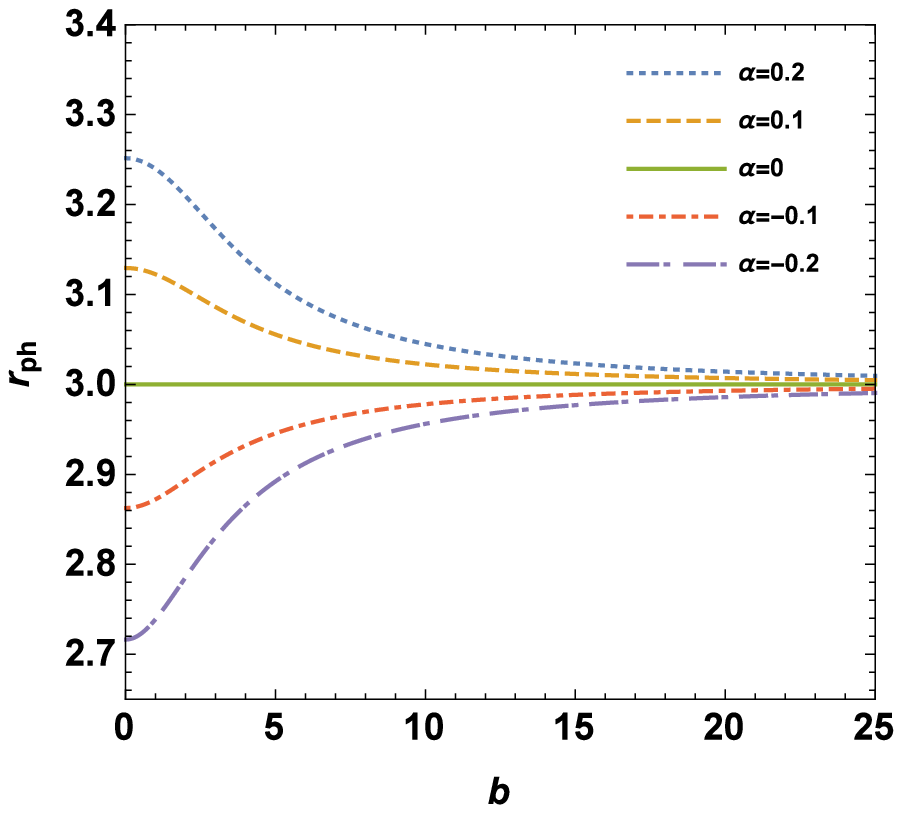}\;\;\;\includegraphics[width=3.8cm]{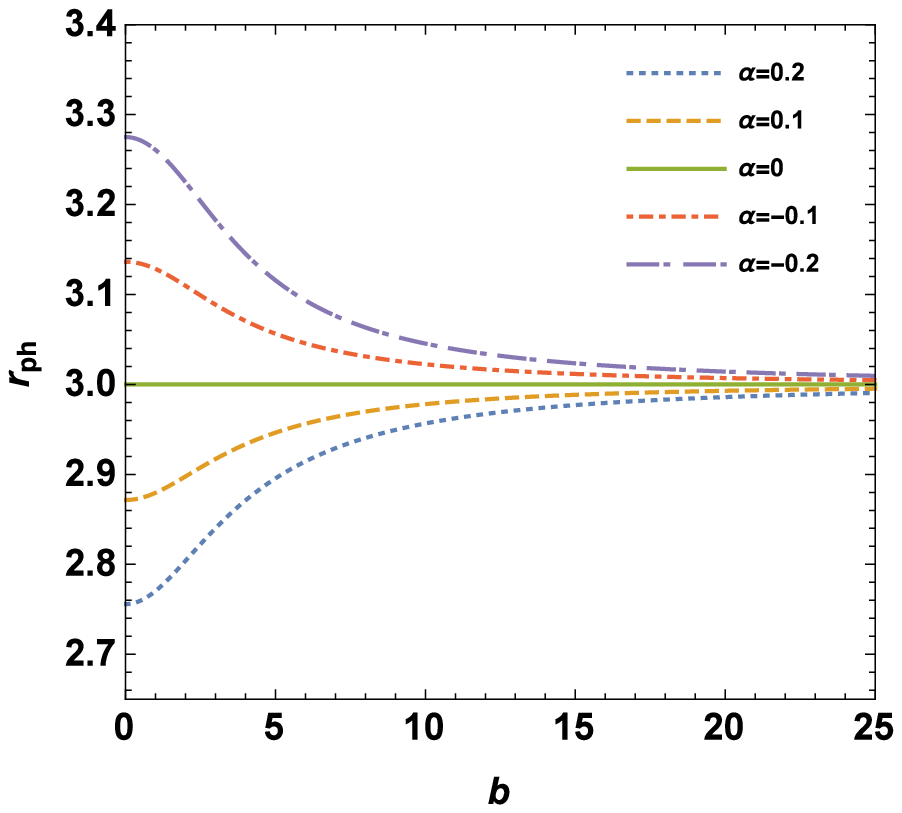}
\caption{Variety of the inner circular orbit radius $r_{ph}$ with
the phantom charge $b$ in the regular phantom black hole
spacetime for fixed $\alpha$. The left and the right are for the photons PPL and PPM,
respectively. Here we set $2M=1$.}
\end{center}
\end{figure}
It is shown that with the increase of the coupling parameter $\alpha$ the inner circular orbit radius $r_{ph}$ for different $b$ increases for PPL and decreases for PPM, which is similar to that in the Schwarzschild black hole spacetime. With the increase of the phantom charge $b$, $r_{ph}$ for the PPL increases for the negative $\alpha$ and decreases for positive one. For the case of PPM, the change of $r_{ph}$ with $b$ is converse to those of PPL. Thus, the inner circular orbit radius $r_{ph}$ depends on the coupling parameter $\alpha$, the phantom charge $b$ and the polarization, which is quite different from that in the case without the coupling in which the inner circular orbit radius $r_{ph}$ is independent of the phantom charge $b$ and the polarization of photon. In other words, the presence of the coupling brings richer behavior for the inner circular orbit radius $r_{ph}$ of the photon.

\section{Double shadows of a regular phantom black hole as photons couple to Weyl tensor}

In the regular phantom black hole spacetime (\ref{metric0}), the light rays from the source can be divided into two classes: the first class can reach the observer at radius coordinate $r_O$ after being deflected by the
black hole. The other class go towards the
horizon of the black hole and are not detected by the observer at $r_O$. It yields a dark region on the observer¡¯s sky called as the shadow of the black hole. In the spherically symmetric spacetime, the shadow of black hole is a dark circular disk and its boundary is determined by light
rays that spiral asymptotically towards a circular light
orbit at radius $r_{ph}$. After some operations, one can find that the angular radius of the
shadow $\alpha_{sh}$  in the regular phantom black hole spacetime (\ref{metric0}) can be modified as
\begin{eqnarray}
\sin^2\alpha_{sh}
=\frac{A(r_{O})C(r_{ph})W(r_{O})}{A(r_{ph})C(r_{O})W(r_{rh})}.\label{shadow}
\end{eqnarray}
The mass of the central object of our Galaxy is evaluated to be $4.4\times 10^6M_{\odot}$  and its distance from the earth is around $8.5kpc$ \cite{grf}, which means the ratio $GM/r_{O} \approx2.4734\times10^{-11}$ .
Combing with Eqs. (\ref{shadow}), we can estimate the values of angular radius of the
shadow yielded by the photon couples with Weyl tensor in a regular phantom spacetime. We present the dependence of the angular radius $\alpha_{sh}$ on the coupling constant $\alpha$ and phantom charge $b$ in Figs.(3) and (4). With the increase of the coupling constant $\alpha$, the angular radius  of the shadow $\alpha_{sh}$ increases for the photon PPL and decrease for the photon PPM. Moreover, with the increase of the phantom charge $b$, $\alpha_{sh}$  increase for two different coupled photons. This means that the size of the shadow of black hole increases with the phantom charge in this case.
\begin{figure}[ht]
\begin{center}
\includegraphics[width=3.8cm]{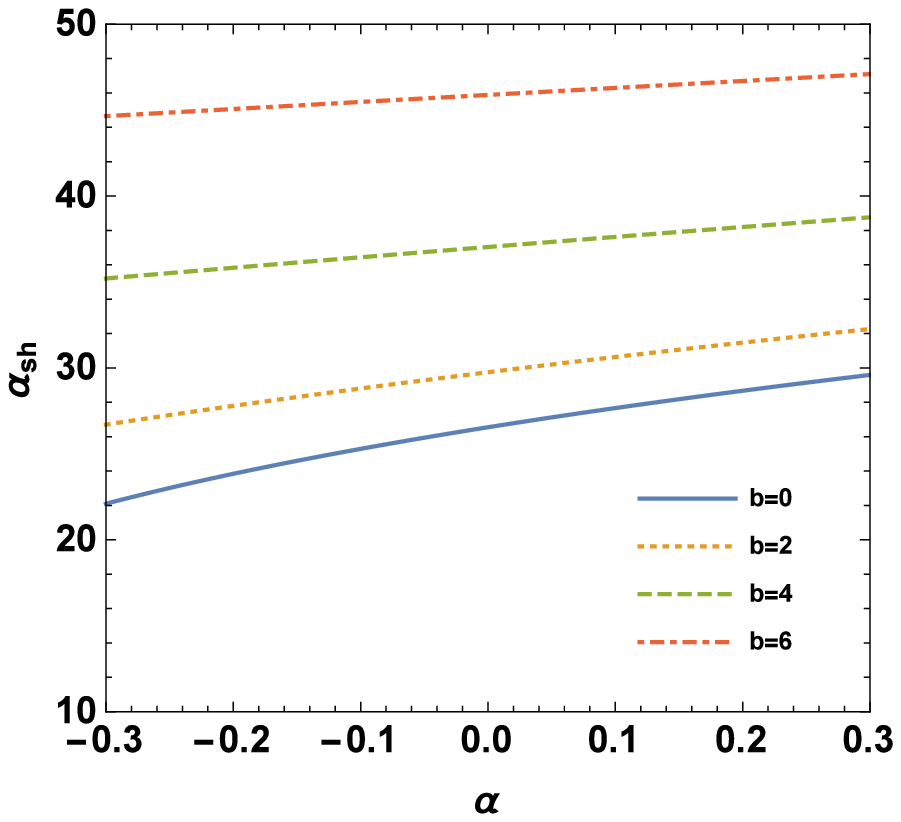}\;\;\;\includegraphics[width=3.8cm]{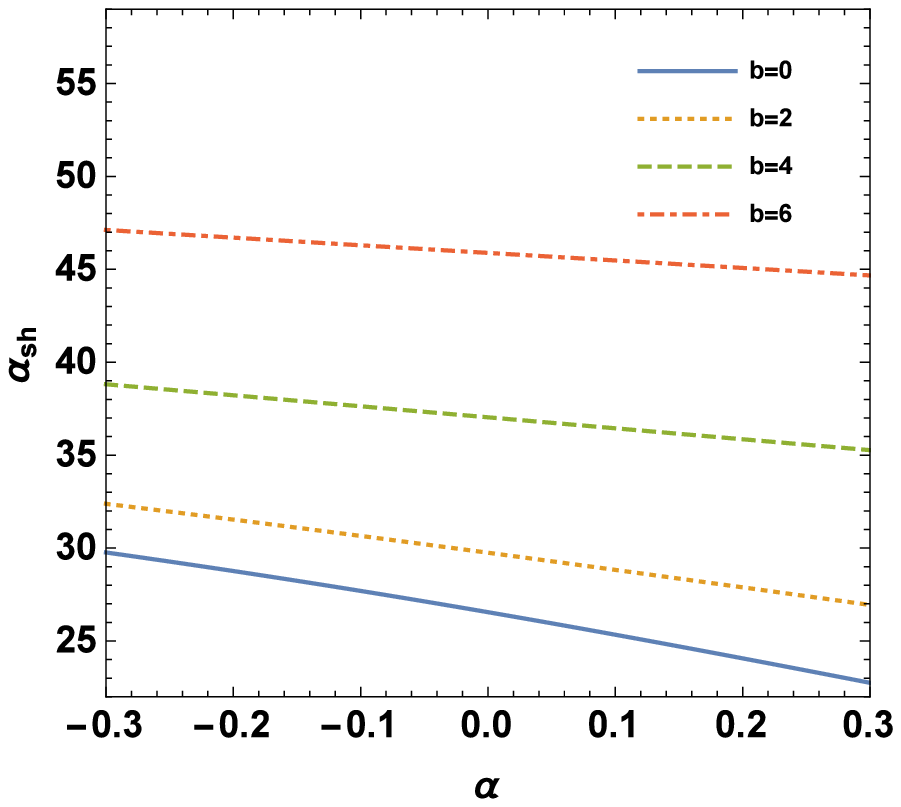}
\caption{Variety of the angular radius of the shadow $\alpha_{sh}$ with
the coupling constant $\alpha$ in the regular phantom black hole
spacetime for fixed $b$. The left and the right are for the photons PPL and PPM,
respectively. Here we set $2M=1$.}
\end{center}
\end{figure}
\begin{figure}
\begin{center}
\includegraphics[width=3.8cm]{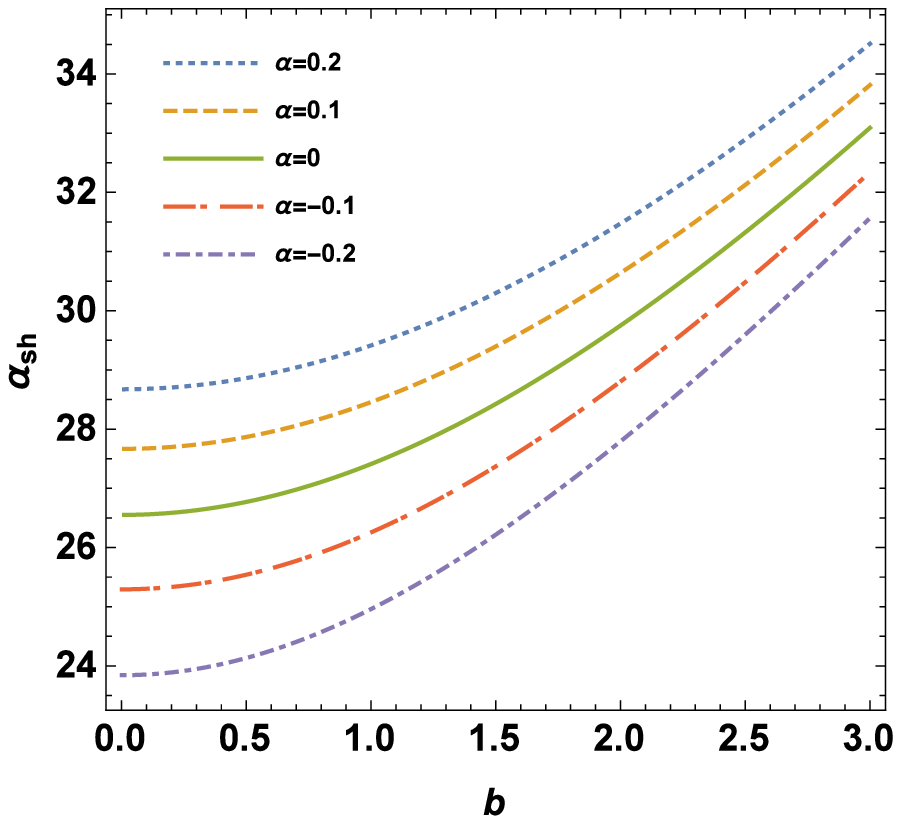}\;\;\;\includegraphics[width=3.8cm]{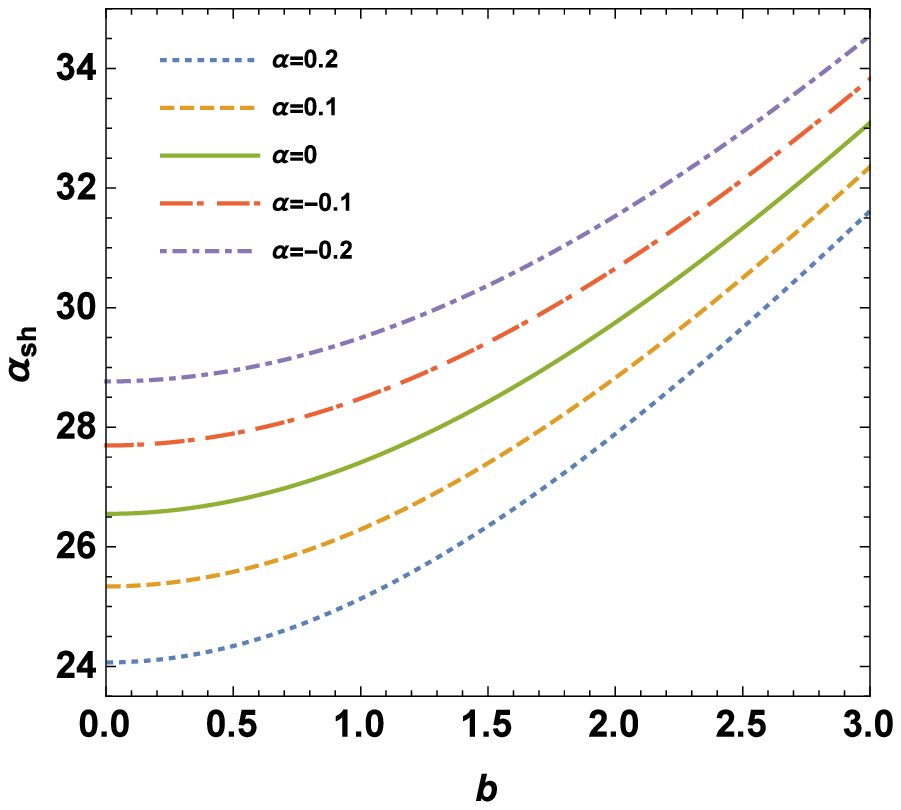}
\caption{Variety of the angular radius of the shadow $\alpha_{sh}$  with
the phantom charge $b$ in the regular phantom black hole spacetime for fixed $\alpha$. The left and the right are for the photons PPL and PPM, respectively. Here we set $2M=1$.}
\end{center}
\end{figure}

Considering that the light radiated from the source could be white light which can be separated into two kinds of linearly polarized light beams, it is naturally expected that there exists double shadow of a regular phantom black hole as photons couple to Weyl tensor. The overlap region of double shadow is called as umbra. From Fig.(3), we find that the region of umbra is determined by PPL as $\alpha $ is negative and by PPM as $\alpha $ is positive. Moreover, one can find that the umbra of black hole increases with the phantom charge and decreases with the coupling strength. In Fig.(5), we studied the dependence of the penumbra of black hole on the phantom charge $b$ and the coupling parameter $\alpha$. It is shown that the angular radius of penumbra of the black hole $\Delta\alpha_{sh}$ decreases with the phantom charge and increases with the coupling strength, which is converse to that of the umbra. As the coupling vanishes, we find that $\Delta\alpha_{sh}=0$, which means that the boundary of shadow caused by PPM is overlapped with that by PPL and then the double shadow of black hole is reduced to a single shadow in this case.
Therefore, the presence of the coupling brings richer behaviors for the shadow of the black hole in the regular phantom black hole spacetime.
\begin{figure}
\begin{center}
\includegraphics[width=3.8cm]{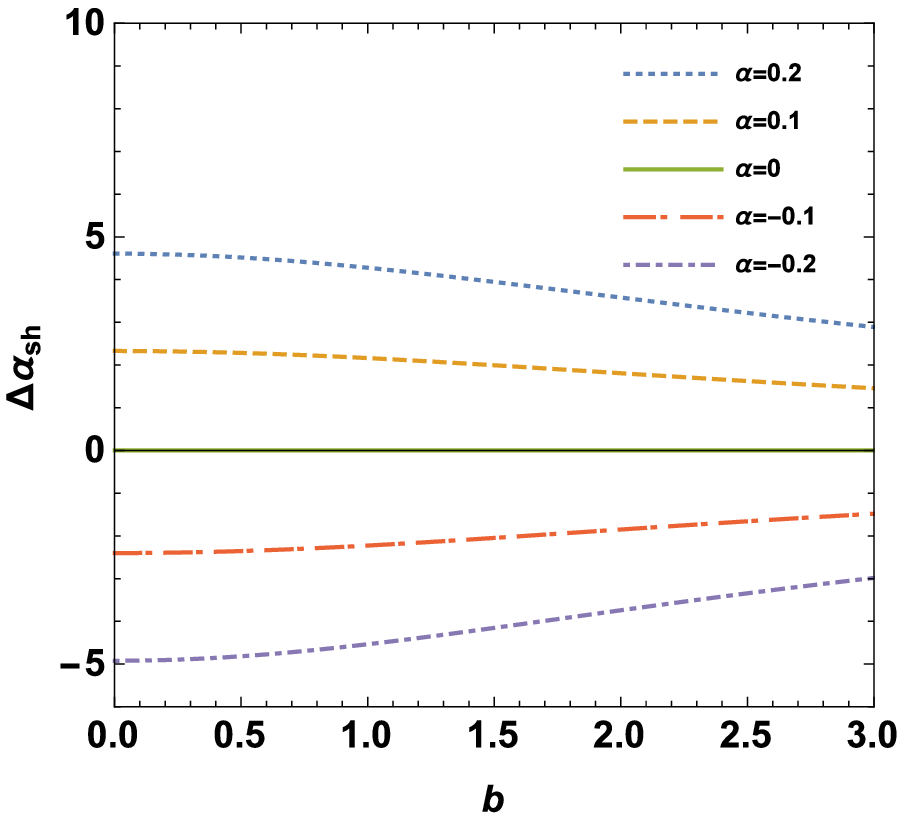}\;\;\;\includegraphics[width=3.8cm]{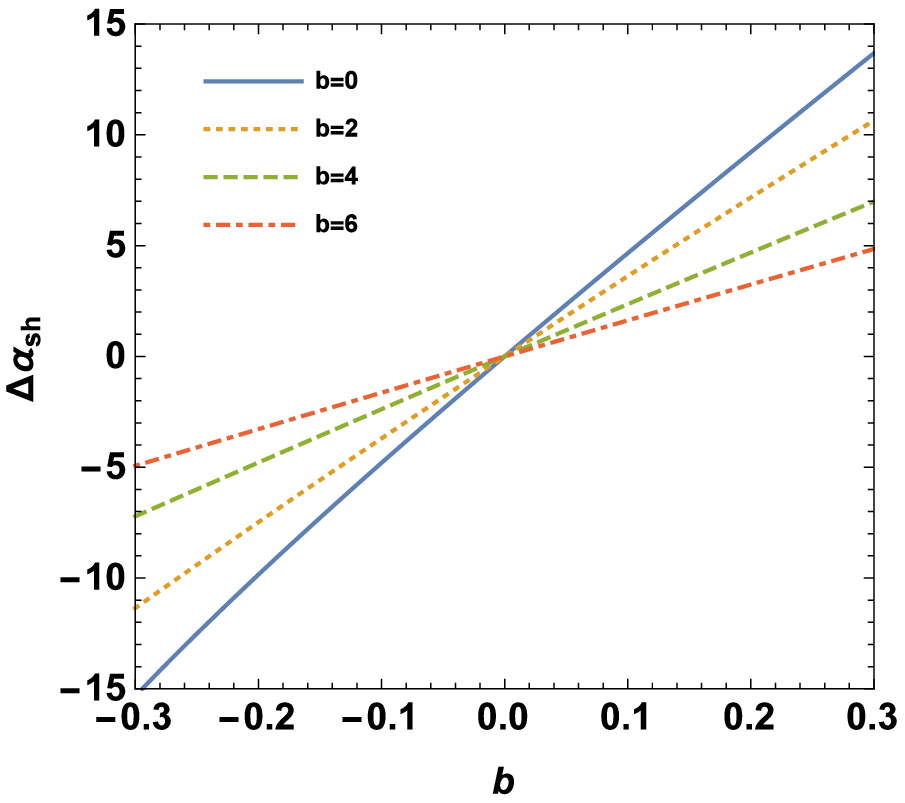}
\caption{Dependence of the penumbra of black hole on
the phantom charge $b$ and the coupling parameter $\alpha$ in the regular phantom black hole spacetime. }
\end{center}
\end{figure}

\section{Summary}

Summary, we have studied the shadow of a regular phantom black hole by photon coupling to Weyl tensor. We find that the propagation of coupled photon and the shadow of the black hole depend sharply on the phantom charge of black hole, the photon polarization directions and the coupling between photon and Weyl tensor. The presence of the coupling results in that the inner circular orbit radius of the coupled photons around black hole depends on the phantom charge and the polarization of photon itself, which is different from that in the non-coupling case in which  the inner circular orbit radius is independent of the phantom charge and the polarization of photon.  With the increase of the phantom charge $b$, the inner circular orbit radius $r_{ph}$ for the PPL increases for the negative $\alpha$ and decreases for positive one. For the case of PPM, the change of $r_{ph}$ with $b$ is converse to those of PPL.

We also studied the double shadow of black hole as photons couple to Weyl tensor, which does not appear in the non-coupling case where only a single shadow emerges. We find that the umbra of black hole increases with the phantom charge and decreases with the coupling strength. The dependence of the penumbra on the phantom charge and the coupling strength is converse to that of the umbra.
Combining with the supermassive central object in our Galaxy, we estimated the shadow of the black hole as the photons couple to Weyl tensor. It would be of interest to generalize our study to other black hole spacetimes, such as Kerr black hole etc. Work in this direction will be reported in the future.

\section{\bf Acknowledgments}
This work was partially supported by the National Natural
Science Foundation of China under Grant No.11275065,  No. 11475061,
the construct program of the National Key Discipline, and the Open
Project Program of State Key Laboratory of Theoretical Physics,
Institute of Theoretical Physics, Chinese Academy of Sciences, China
(No.Y5KF161CJ1).

\end{document}